# Strain-Induced Water Dissociation on Supported Ultrathin Oxide Films

Zhenjun Song, Jing Fan, and Hu Xu∗

Department of Physics, South University of Science and Technology of China, Shenzhen, 518055, China

Controlling the dissociation of single water molecule on an insulating surface plays a crucial role in many catalytic reactions. In this Letter, we have identified the enhanced chemical reactivity of ultrathin MgO(100) films deposited on Mo(100) substrate that causes water dissociation. We reveal that the ability to split water on insulating surface closely depends on the lattice mismatch between ultrathin films and the underlying substrate, and substrate-induced in-plane tensile strain dramatically results in water dissociation on MgO(100). Three dissociative adsorption configurations of water with lower energy are predicted, and the structural transition going from molecular form to dissociative form is almost barrierless. Our results provide an effective avenue to achieve water dissociation at the single-molecule level and shed light on how to tune the chemical reactions of insulating surfaces by choosing the suitable substrates.

The interaction of water with metal oxide surfaces has attracted considerable interest due to their important promising applications in photocatalysis, electrochemistry, and sensors[1,2]. Understanding the mechanism of water dissociation on oxide surfaces is of fundamental interest to uncover how chemical reactions work involving water dissociation. More importantly, if we know how to control the adsorption states of water then we can selectively tune chemical reactions. Usually, hydrogen bonds play an important role in describing the structural geometry of partial dissociation of water on oxide surfaces[3,4], while the intrinsic surface states are the driving force to induce water dissociation on metal oxide surfaces at various coverages[5], especially at lower coverage.

Among oxides, MgO(100) is a good model system due to its simple structural and electronic properties to reveal chemical reactivity and catalytic activity of metal oxides. Water adsorption on MgO(100) surfaces has been intensively studied for many years, and it is well known that water will partially dissociate on MgO(100) surface at higher coverage due to strong inter-molecular hydrogen bonding[6,7], while water prefers to adsorb in molecular form at lower coverage[7]. As

MgO(100) is one of typical insulating surfaces, it is inactive in surface reactions and usually chemically inert towards $O_2$, $H_2O$, and other molecules. Recently, ultrathin MgO films deposited on metal substrates have received extensively studied[8-15] due to their potential applications in catalysts. The substrate-induced enhancement of chemical reactivity has been widely reported[8,13,15]. For example, it is revealed that $O_2$ can be activated to form an $O_2\cdot^-$ radical on MgO(100)/Mo(100) surface[15]. In addition, water adsorption on MgO(100)/Ag(100) has been studied recently, and the energy barriers for water dissociation have been effectively reduced by tuning film thickness[10], introducing interface defects[14] or 3d transition metal dopants[11]. Furthermore, energy differences between molecular and dissociative adsorption of water on MgO(100)/Ag(100) also decrease compared with the case for stoichiometric MgO(100) surface[10]. Unfortunately, an intact water molecule is still energetically favorable on MgO(100)/Ag(100) surface[10,11,14]. Although many efforts have been made, it remains challenging to achieve one single water molecule dissociation on MgO(100) surfaces. Therefore, it is still desirable and significant to make further efforts to strengthen the chemical activity of MgO(100) to split water.

In this Letter, the strain-induced water dissociation on MgO(100) is proposed theoretically, and the mechanism of water dissociation on supported ultrathin MgO(100) films is also uncovered. We demonstrate that the dissociation of a single water molecule on Mo-suppported ultra-thin MgO(100) films is exothermal, and the activation barriers from molecular adsorption state of water to dissociative adsorption can be reduced significantly (nearly zero). More importantly, we provide a feasible way to modulate the adsorption states of water on supported insulating surfaces.

Density-functional theory (DFT) calculations have been performed using Vienna ab initio simulation pack-age (VASP)[16,17] to study the water adsorption behaviors. Perdew-Burke-Ernzerhof (PBE) functional[18] within generalized gradient approximation (GGA) is chose to describe exchange and correlation effects, as PBE functional gives the excellent description of hydrogen bonds[19]. Projector augmented wave (PAW) method[20] is used to describe the interactions between valence and core electrons. The energy cutoff is 500 eV, and the convergence criterion on each atom during structural relaxations is less than 0.02 eV/Å. In order to avoid the inter-molecular interaction we present results using a p(4×4) Mo(100) surface, where the distance between the adjacent water molecules is 12.60 Å. Four atomic Mo layers are used to mimic the substrate, and the bottom two layers are fixed at their bulk positions while the top two layers are allowed to relax. One to five monolayers (ML) of MgO(100) are adopted as the ultrathin MgO films. A vacuum region of 15 Å is introduced to separate the neighbouring slabs. The (2×2×1) and (4×4×1) k-point Monkhorst-Pack samplings[21] are used for structural relaxations and total energy calculations, respectively. The energy barriers and transition states are estimated by using the climbing image nudged elastic band (CI-NEB) method[22].

The calculated lattice constants for body-centered cubic (bcc) Mo and rock-salt MgO bulk are respective 3.15 Å and 4.21 Å, which are in good agreement with the experimental values[23]. Owing to the small mismatch between MgO(100) and Mo(100) surfaces, it is usually to use Mo(100) as the substrate to study ultrathin MgO(100) films. The lattice mismatch between MgO(100) and Mo(100) is 5.1%, therefore MgO ultrathin films supported on Mo(100) will slightly expand compared with their bulk position. The interlayer distance between Mo substrate and 1 ML MgO(100) is 2.10 Å, while this distance increases to 2.15 Å for 2-5 ML MgO(100). Oxygen atoms at the interface prefer to bond to surface Mo atoms, which is in line with prior results[24].

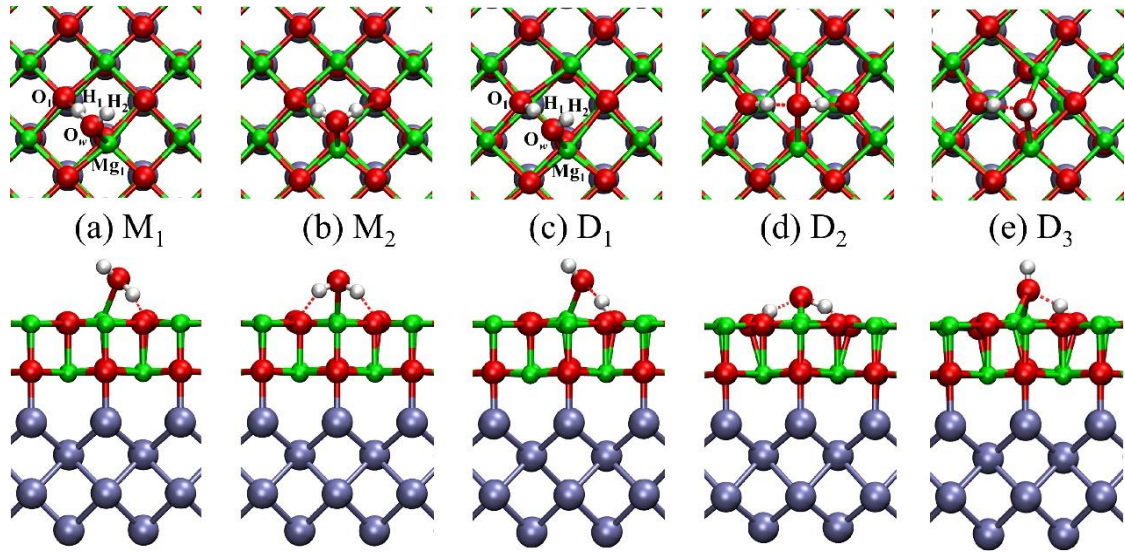

FIG. 1. (Color online) The top and side views of water adsorption on Mo(100) supported MgO(100) surfaces. Water adsorbs in molecular form with (a) one or (b) two hydrogen bonds between water and surface oxygen. (c-e) Water adsorbs in dissociative form.

It is well known that water molecule prefers to ad-sorb on the stoichiometric MgO(100) surface in molecular form at low coverage[7]. Then it will form two nearly degenerate adsorption structures with one or two hydrogen bonds between water and surface oxygen, and the corresponding adsorption energies per water are around -0.45 eV. We then study water behaviors on MgO(1 - 5

ML)/Mo(001) surfaces. Water will initially lands on MgO(001)/Mo(001) surfaces in the molecular form. Similarly, it is also found that water molecules have two possible adsorption configurations in molecular form with nearly degenerate adsorption energy. One molecular configuration ($M_1$) is that there is one strong hydrogen bond between water and surface oxygen with the distance of 1.38 Å, (see Fig. 1(a)), while another molecular adsorption ($M_2$) has two identical weak hydrogen bonds with the distance of around 1.68 Å(see Fig. 1(b)). The adsorption energies per water for both M1 and M2 on MgO(1-5 ML)/Mo(100) are from -0.67 eV to -0.75 eV, while the adsorption energies per water on ultrathin MgO(100) films deposited on Ag(100) are around -0.5 eV. The results indicate that molecular adsorption of water can be significantly strengthened by the Mo(100) substrate. In addition, the adsorption energy per water are almost insensitive to film thickness.

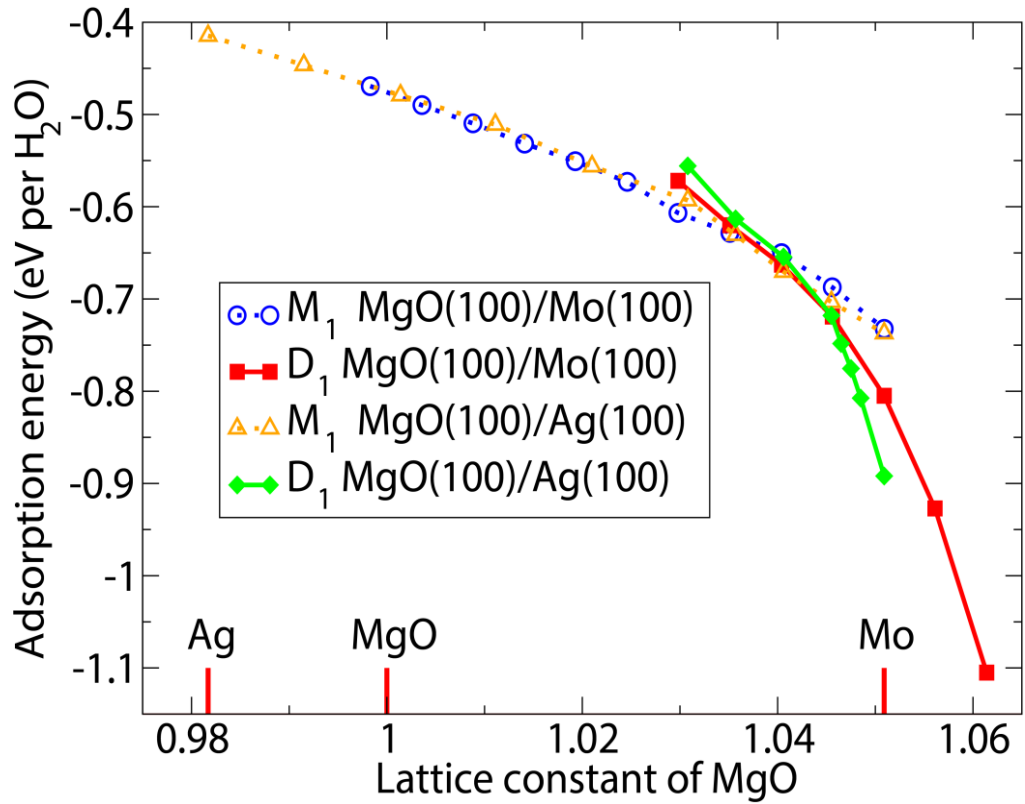

FIG. 2. (Color online) Adsorption energy per water as a function of MgO lattice constant on Ag- or Mo-supported 2 ML MgO(100) thin films. The optimized lattice constants of Ag and Mo are also marked relative to MgO.

The questions is where the adsorption energy differences for water adsorption on ultrathin MgO(100) films deposited on different metal substrates come from. It is clearly that MgO lattice is slightly contracted by 1.8% on Ag while expanded by 5.1% on Mo substrate. Is MgO lattice expansion induced by Mo substrate responsible for the enhancement of water adsorption? To verify our assumption, we have calculated the adsorption energy per water as a function of MgO lattice on MgO(2 ML)/Ag(100) and MgO(2 ML)/Mo(100) surfaces shown in Fig. 2. When ultrathin MgO(100) films deposited on Ag(100) substrate, the lattice of MgO will be shortened by 1.8%, and the corresponding adsorption energy per water is -0.41 eV. While the adsorption energy per water is -0.74 eV on Mo-supported MgO(100). If we keep the lattice parameters of MgO(100)/Mo(100) unchanged, and just replace Mo by Ag, then in this case the adsorption energy per water is -0.73 eV. Our results indicate that the adsorption energy for molecular adsorption almost linearly increases with the increasing of MgO lattice constant. In other words, the adsorption energy closely depends on the lattice parameter of MgO, while charge effect does not play an important role in water dissociation. The results definitely indicate that the expansion of MgO lattice will remarkably strengthen the interaction of water with MgO(100) surface. This is because the increment of the bond length of MgO will reduce bond strength significantly, resulting in the enhancement of their reactivity for water splitting.

Now that the interaction of water with ultrathin MgO(100) films has been greatly improved by Mo(100) substrate, it is interesting to study whether MgO(100)/Mo(100) is reactive for water dissociation. In contrast to adsorption behaviors of water on MgO(100)/Ag(100), water will easily dissociate on MgO(100)/Mo(100) surface, which implies that the ability to split water on ultrathin MgO(100) films is notably improved by Mo(100) substrate. Three possible dissociative configurations $D_1$, $D_2$, and $D_3$ are shown in Figs. 1(c)-1(e), respectively. The adsorption energies for molecular and dissociative adsorption for MgO(1-5 ML)/Mo(100) are listed in Table I. From Table I, we can find that the dissociative configurations are favored over molecular adsorption.

To uncover the dissociative mechanism of water, we systematically study the structural configurations of $M_1$ and $D_1$ using MgO(2 ML)/Mo(100) surface. The corresponding structural parameters and adsorption energy per water as a function of MgO lattice are listed in Table II. The MgO lattice increases gradually from +0.0% to 5.1%, where MgO lattice with 5.1% expansion is equal to that of Mo lattice. The results clearly show that the bond length of $O_w$-$H_1$ in water steadily increases from 1.02 to 1.12 Å along with MgO lattice expansion range from 0.0% to 5.1%, where the bond length of $O_w$-$H_2$ in water is unaffected by the change of MgO lattice. Accordingly, the hydrogen bond between water and surface oxygen ($O_1$-$H_1$) gets shorter by 0.29 Å. The bond length elongation of $O_w$-$H_1$ and the shortening of $O_1$-$H_1$ indicate that water molecule tends to dissociate. In addition, from Table II we can clearly note that the bond length of O -$Mg_1$ decreases significantly with the increase of unit cell size, which implies the stronger interaction between water and surface. Furthermore, the bond length of $O_1$-$Mg_1$ increases by around 0.5 Å with the induced strain by Mo substrate. The angle of $O_1$-$Mg_1$-$O_w$ ($\theta$) also de-creases by 10˚.

As shown in Fig. 2, the slopes of adsorption energy for molecular and dissociative water behave differently. The dissociated water has a steeper slope than that of molecular one, as a result water prefers to dissociate on the MgO(100) surface when 4% interfacial strain is applied. As we know that the interfacial strain will change the lattice of ultrathin MgO films as the lattice constants of metal substrates vary. When ultrathin MgO films deposited on Mo(100) substrate, the MgO lattice is enlarged by 5.1%, so water prefers to dissociate on MgO(100)/Mo(100) surface. While MgO lattice shrinks 1.8% constrained on the Ag(100) substrate, thus water does not prefer to dissociate on this system. In fact, if we assume that Ag has the same lattice as Mo, water will also dissociate on MgO(100)/Ag(100) surface (see Fig. 2). In addition to metal substrate, the thickness of MgO(100) films also have some influence on the dissociative adsorption energy of water. For example, we can find that water in dissociative form on MgO(1-2 ML)/Mo(100) has much lower adsorption energy than that on MgO(3-5 ML)/Mo(100).

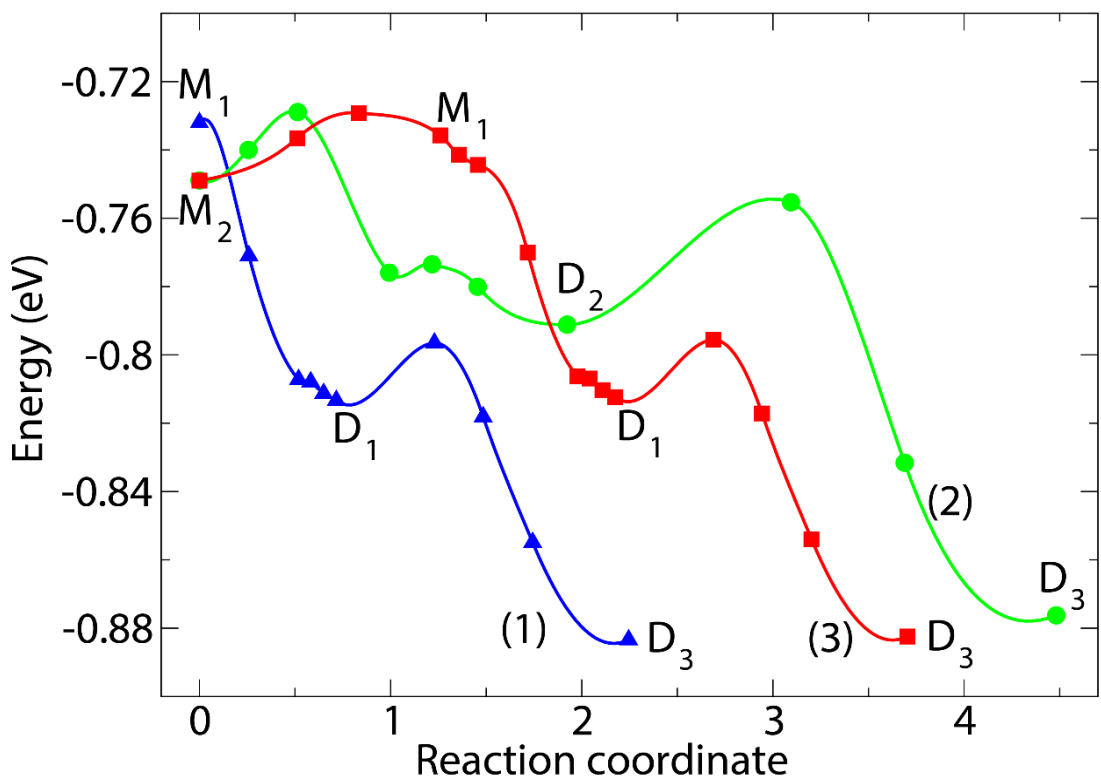

FIG. 3. (Color online) Three possible potential-energy profiles for molecular to dissociative adsorption of water on MgO(2ML)/Mo(100): (1) M1→D1→D3, (2) M2→D2→D3, and (3) M2→M1→D1→D3.

The corresponding reaction pathways for water dissociation on MgO(2 ML)/Mo(100) surfaces are shown in Fig. 3. Depending on the water adsorption configurations on MgO(2 ML)/Mo(100) surface, there may exist three possible water dissociation channels. For channel one (see blue line of Fig. 3), $M_1$ will spontaneously transfer to $D_1$ passing through a barrierless pathway with the energy gain of 0.08 eV. Then $D_1$ can easily transfer to $D_3$ by climbing over a small barrier of 0.02 eV. $D_3$ is the most energetically favorable adsorption configuration with the lowest dissociative adsorption energy of -0.88 eV. For $D_3$, the $O_wH$ group binds to two surface Mg atoms forming two strong bonds. In addition, there exists one strong hydrogen bond between $O_w$ of the dissociated water and hydrogen binding to surface oxygen. Another dissociation channel (see green line in Fig. 3) is from $M_2$ to $D_3$ via $D_2$. It needs to overcome a very small barrier of 0.02 eV for water to dissociate initially, then it will form the meta-stable dissociative configuration of $D_2$. There are two hydrogen bonds for $D_2$. One hydrogen bond is that the dissociated H points to dissociated $O_wH$ and another one forms between H from dissociated $O_wH$ and surface oxygen. Afterwards, $D_3$ also forms by striding over the energy barrier of 0.04 eV. Furthermore, $M_2$ may transfer to $M_1$ due to the small reaction barrier of 0.02 eV, then $D_3$ forms going across $D_1$, which is the third dissociation channel (see red line in Fig. 3). As energy barriers during water dissociation are relatively low for all the dissociation channels, there may exist multiple dissociation pathways for water on MgO(100)/Mo(100) surface. Among these, the channel one should be the most likely channel for water dissociation.

In summary, we have performed a systematic study to investigate the interaction of water with Mo-supported ultrathin MgO(100) films. The understanding of how water interacts with metal oxide surfaces is important in uncovering the interfacial phenomena. The single water molecule has been successfully split on insulating surface by choosing the suitable metal substrate. The mechanism of water dissociation on MgO(100)/Mo(100) surface has been revealed. The interfacial tensile strain due to lat-tice mismatch will cause the expansion of MgO lattice, and 4% expansion of MgO lattice will result in the dis-sociation of water on supported MgO(100) surface. Our results provide an effective method to enhance the surface reactivity towards water by choosing the suitable substrate.

**Acknowledgement**

This work is supported by the National Natural Sci-ence Foundation of China (NSFC, Grant Nos. 11204185 and 11334003) and internal Research Grant Program (FRG-SUSTC1501A-35).

1   Diebold, U. The surface science of titanium dioxide. *Surf Sci Rep* **48**, 53-229, doi:Pii S0167-5729(02)00100-0 Doi 10.1016/S0167-5729(02)00100-0 (2003).
2   Verdaguer, A., Sacha, G. M., Bluhm, H. & Salmeron, M. Molecular structure of water at interfaces: Wetting at the nanometer scale. *Chem Rev* **106**, 1478-1510, doi:10.1021/cr040376l (2006).
3   Dulub, O., Meyer, B. & Diebold, U. Observation of the dynamical change in a water monolayer adsorbed on a ZnO surface. *Phys Rev Lett* **95**, doi:Artn 136101 10.1103/Physrevlett.95.136101 (2005).
4   Ferry, D. *et al.* Observation of the second ordered phase of water on the MgO(100) surface: Low energy electron diffraction and helium atom scattering studies. *J Chem Phys* **105**, 1697-1701, doi:Doi 10.1063/1.472028 (1996).
5   Xu, H. *et al.* Splitting Water on Metal Oxide Surfaces. *J Phys Chem C* **115**, 19710-19715, doi:10.1021/jp2032884 (2011).
6   Giordano, L., Goniakowski, J. & Suzanne, J. Partial dissociation of water molecules in the (3 x 2) water monolayer deposited on the MgO (100) surface. *Phys Rev Lett* **81**, 1271-1273, doi:DOI 10.1103/PhysRevLett.81.1271 (1998).
7   Cho, J. H., Park, J. M. & Kim, K. S. Influence of intermolecular hydrogen bonding on water dissociation at the MgO(001) surface. *Phys Rev B* **62**, 9981-9984, doi:DOI 10.1103/PhysRevB.62.9981 (2000).

8   Hellman, A., Klacar, S. & Gronbeck, H. Low Temperature CO Oxidation over Supported Ultrathin MgO Films. *J Am Chem Soc* **131**, 16636-+, doi:10.1021/ja906865f (2009).

9   Shin, H. J. *et al.* State-selective dissociation of a single water molecule on an ultrathin MgO film. *Nat Mater* **9**, 442-447, doi:10.1038/NMAT2740 (2010).

10  Jung, J., Shin, H. J., Kim, Y. & Kawai, M. Controlling water dissociation on an ultrathin MgO film by tuning film thickness. *Phys Rev B* **82**, doi:Artn 085413 10.1103/Physrevb.82.085413 (2010).

11  Jung, J., Shin, H. J., Kim, Y. & Kawai, M. Ligand Field Effect at Oxiide-Metal Interface on the Chemical Reactivity of Ultrathin Oxide Film Surface. *J Am Chem Soc* **134**, 10554-10561, doi:10.1021/ja302949j (2012).

12  Honkala, K., Hellman, A. & Gronbeck, H. Water Dissociation on MgO/Ag(100): Support Induced Stabilization or Electron Pairing? *J Phys Chem C* **114**, 7070-7075, doi:10.1021/jp9116062 (2010).

13  Savio, L., Celasco, E., Vattuone, L. & Rocca, M. Enhanced reactivity at metal-oxide interface: Water interaction with MgO ultrathin films. *J Phys Chem B* **108**, 7771-7778, doi:Doi 10.1021/Jp0360873 (2004).

14  Jung, J., Shin, H. J., Kim, Y. & Kawai, M. Activation of Ultrathin Oxide Films for Chemical Reaction by Interface Defects. *J Am Chem Soc* **133**, 6142-6145, doi:10.1021/ja200854g (2011).

15  Gonchar, A. *et al.* Activation of Oxygen on MgO: O-2(center dot-) Radical Ion Formation on Thin, Metal-Supported MgO(001) Films. *Angew Chem Int Edit* **50**, 2635-2638, doi:10.1002/anie.201005729 (2011).

16  Kresse, G. & Hafner, J. Abinitio Molecular-Dynamics for Liquid-Metals. *Phys Rev B* **47**, 558-561, doi:DOI 10.1103/PhysRevB.47.558 (1993).

17  Kresse, G. & Furthmuller, J. Efficient iterative schemes for ab initio total-energy calculations using a plane-wave basis set. *Phys Rev B* **54**, 11169-11186, doi:DOI 10.1103/PhysRevB.54.11169 (1996).

18  Perdew, J. P., Burke, K. & Ernzerhof, M. Generalized gradient approximation made simple. *Phys Rev Lett* **77**, 3865-3868, doi:DOI 10.1103/PhysRevLett.77.3865 (1996).

19  Ireta, J., Neugebauer, J. & Scheffler, M. On the accuracy of DFT for describing hydrogen bonds: Dependence on the bond directionality. *J Phys Chem A* **108**, 5692-5698, doi:10.1021/jp0377073 (2004).

20  Blochl, P. E. Projector Augmented-Wave Method. *Phys Rev B* **50**, 17953-17979, doi:DOI 10.1103/PhysRevB.50.17953 (1994).

21  Monkhorst, H. J. & Pack, J. D. Special Points for Brillouin-Zone Integrations. *Phys Rev B* **13**, 5188-5192, doi:DOI 10.1103/PhysRevB.13.5188 (1976).

22  Henkelman, G., Uberuaga, B. P. & Jonsson, H. A climbing image nudged elastic band method for finding saddle points and minimum energy paths. *J Chem Phys* **113**, 9901-9904, doi:Pii [S0021-9606(00)71246-3] Doi 10.1063/1.1329672 (2000).

23  Prada, S., Martinez, U. & Pacchioni, G. Work function changes induced by deposition of ultrathin dielectric films on metals: A theoretical analysis. *Phys Rev B* **78**, doi:Artn 235423 10.1103/Physrevb.78.235423 (2008).

24  Giordano, L., Baistrocchi, M. & Pacchioni, G. Bonding of Pd, Ag, and Au atoms on MgO(100) surfaces and MgO/Mo(100) ultra-thin films: A comparative DFT study. *Phys Rev B* **72**, doi:Artn 11540310.1103/Physrevb.72.115403 (2005).